\documentstyle[12pt]{article}

\begin{document}

\baselineskip=24pt

\begin{center}
{\Large \bf MARKOV FORM OF THE METHOD OF NONEQUILIBRIUM STATISTICAL
OPERATOR}
\end{center}

\medskip

\begin{center}
{M.D.Zviadadze, A.G.Kvirikadze, M.M.Sozashvili, L.O.Tkeshelashvili} \\
\end{center}

\medskip

\begin{center}
{\it Institute of Physics, Georgian Academy of Sciences, \\
6 Tamarashvili str., GE-380077 Tbilisi, Georgia}\\
E-mail: nic@physics.iberiapac.ge
\end {center}

\begin{abstract}
\baselineskip=24pt

The general principles of the choice of the reduced description parameters
of nonequilibrium states $\gamma_{\alpha}(t)$ and the construction of the
nonequilibrium statistical operator (NSO) $\rho (t)$ are discussed.
On the basis of
Kavasaki - Gunton projection operator an explicit non-Markov form of NSO
with respect
to $\gamma_{\alpha}(t)$ parameters is obtained and the equivalence
of the method to the other methods using boundary conditions for
$\rho(t)$ or "mixing" operator is shown. An exact NSO transformation to
Markov form with respect to $\gamma_{\alpha}(t)$ at arbitrary dependence of
$V_t$ perturbation on time is made. The generalized kinetic equations not
containing the memory with respect to $\gamma_{\alpha}(t)$ are obtained being
convenient for practical application.

\end{abstract}
\newpage
\baselineskip=24pt

\begin{center}
{1. Introduction}
\end{center}

The existing method of theoretical study of irreversible processes [1-3]
are based on N.N.Bogoliubov concept [1] of sequential reductions of
parameter number necessary for description of nonequilibrium states of
macroscopic system in the process of its evolution in time. In accordance
with this concept a quasiequilibrium states is determined
in the system at $t \gg \tau_0$ ($\tau_0$ - is the chaotization time)
that can be characterized by giving average values
$\gamma_{\alpha}(t) = Sp\{\rho (t)\hat {\gamma_{\alpha}}\}$
varying comparatively slowly in time and generally speaking in space.
$\{\gamma_{\alpha}\}$ is a set of limited linearly independent
operators ($\alpha$ is the complex of discrete and continuous indices on
which $\hat {\gamma_{\alpha}}$ operators depend). $\gamma_{\alpha}(t)$
is generally called the generalized thermodynamical coordinates (GTC)
of system. Nonequilibrium statistical operator (NSO) $\rho (t)$
giving a rough description of the system at $t \gg \tau_0$ is,
in the general case, a functional depending on operators
$\hat {\gamma_{\alpha}}$ and  GTC $\gamma_{\alpha}(t)$.
$\gamma_{\alpha}(t)$ functions are determined by solution of the closed
system of generalized kinetic equations (GKE) obtained by averaging
of operator equations of motion for $\hat {\gamma_{\alpha}}$
over NSO $\rho (t)$.

The choice of $\hat {\gamma_{\alpha}}$ operators should be made individually
for each specific system taking into account the hierarchy of the
interactions existing in it.

\begin{center}
{2. Choice of GTC}
\end{center}

Let us consider space-uniform nonequilibrium system with time-dependent
Hamiltonian $H_t$:

$$
H_t = H_0 + V_t; \qquad V_t \equiv V + H_{F}(t) \ll
H_0
\eqno(2.1)
$$
$H_0 -$ is the main Hamiltonian defining a set of operators
$\{\hat {\gamma_{\alpha}}\}$ and  $V_t$ contains relatively
weak interaction including those with variable external fields
(addend $H_{F}(t)$) that do not affect the choice of operators
$\hat {\gamma_{\alpha}}$.

Generally, the degree of freedom of macroscopic system can be divided
into separate quasi-independent groups called subsystems. Then, $H_0$
is the Hamiltonian of free subsystems including as a independent subsystems
the effective multipart interaction $H_{eff}$ responsible for
chaotization time $\tau_0$; $V $ is the interaction between subsystems.
Such situation is realized in the problems of magnetic resonance in
solids [4].

At investigation of the kinetics of condensed medium on the basis of the
idea of quasiparticles [5],  as well as in gases [1] $H_0$ is the
Hamiltonian of free quasiparticles and the interaction
between them is described by operator $V$. In this case $\tau_0$
is of the order of collision time and does not depend neither on
$H_0$, nor on $V$ value.

The main requirement for $\hat {\gamma_{\alpha}}$,
operators is [3]:
$$
iL_0\hat {\gamma_{\alpha}} \equiv \frac {1}{i\hbar} [\hat {\gamma_{\alpha}},\\
H_0] = i \sum_{\beta} \alpha_{\alpha \beta} \hat {\gamma_{\beta}}
\eqno(2.2)
$$
$\alpha_{\alpha \beta}$ is the matrix of $c$ numbers defined by
properties of $H_0$ operator symmetry. For solid spin systems the
following requirement should be added to (2.2)

$$
[\hat {\gamma_{\alpha}}, \quad  H_{eff}] = 0.
\eqno(2.3)
$$

The idea of (2.2) and (2.3) conditions is that the set of
$\{\hat {\gamma_{\alpha}}\}$ is formed by the main Hamiltonian
$H_0$ independent of $V_t$ perturbation. Operators $\hat {\gamma_{\alpha}}$
while commutating with $H_0$ are to be closed on themselves,  i.e.
they should commutate with the effective interaction when it is included
into $H_0$, but not necessarily commutate between themselves.

According to (2.1) - (2.3) the evolution of macroscopic
parameters  $\gamma_{\alpha}(t)$ in time includes the dynamic
change at the expense of $H_0$ (being considered exactly)
and  $H_F(t) \ll H_0$ (being considered according to the perturbation
theory), as well as the relaxation change at the expense of
 $V \ll H_0$ with relaxation time $\tau_r$, and in accordance with
the principles of
reduced description  $\tau_r \gg \tau_0$.

\begin{center}
{3. The method of construction of nonequilibrium statistical operator}
\end{center}

The simplest method of NSO construction is the application of Liuville
equation with infinitively small source [6]:

$$
\{\frac{\partial }{\partial t} + iL(t) \}\rho (t) =
- \epsilon \{\rho (t) - P(t) \rho (t) \}; \qquad \epsilon \rightarrow +0;
\eqno(3.1)
$$
where Liuville operators are determined by the expression:

$$
L(t) = L_0 + L_{v}(t);$$
$$iL_{0}A = \frac {1}{i\hbar}[A,H_0];$$
$$ iL_{v}(t)A = \frac {1}{i\hbar}[A, V_t].
$$

Projection operator $P(t)$ determines the degree of rough description of
nonequilibrium states of the system and the structure of NSO
 $\rho(t)$ at $t \gg \tau_0$. It is convenient to use Kavasaki - Gunton
projection operator as a $P(t)$ [7]:

$$
P(t)A = \rho_{q}(t)SpA +
\sum_{\alpha}\frac{\partial \rho_{q}(t)}
{\partial \gamma_{\alpha}(t)} \{Sp (\hat {\gamma_{\alpha}}A) - \\
\gamma_{\alpha}(t)SpA\}; \eqno(3.2)
$$
where
$$
\rho_q (t) = Q^{-1}_q exp\{-\sum_{\alpha} F_{\alpha}(t)
\hat {\gamma_{\alpha}}\};$$
$$Q_q = Sp \exp\{-\sum_{\alpha}F_{\alpha}(t)\hat {\gamma_{\alpha}}\};
\eqno(3.3)
$$
$\rho_{q}(t)$ is the quasiequilibrium statistical operator depending
on $t$ implicitly through GTC $\gamma_{\alpha}(t)$. The functional connection
$F_{\alpha}(t) \equiv F_{\alpha}(\gamma_{\alpha}(t))$ is determined
by matching conditions

$$
\gamma_{\alpha}(t) = Sp\{\rho (t) \hat{\gamma_{\alpha}}\} = \\
Sp\{\rho_q(t)\hat{\gamma_{\alpha}}\}, \eqno(3.4)
$$
providing the coincidence of true and quasiequilibrium values of
macroscopic parameters and the conjugation of two sets of
$\gamma_{\alpha}(t)$ and $F_{\alpha}(t)$ functions in the sense of
nonequilibrium thermodynamics [2].

According to  (3.2) and (3.4) $P(t)$ has the following properties:

$$
P(t)\rho (t) = \rho_q (t); $$
$$P(t)\frac {\partial\rho (t)}{\partial t} = \frac{\partial \rho_{q}(t)}
{\partial t}; $$
$$P(t^{\prime})P(t)A = P(t^{\prime})A. \eqno(3.5)
$$

Eq. (3.1) is similar to usual Liuville equation:
$$
\{\frac{\partial}{\partial t} + iL(t)\}\rho (t) = 0 \eqno(3.6)
$$
and to the boundary condition of delay type [6,8]
$$
T exp \{i \int_{t}^{t+\tau} L(t^{\prime})dt^{\prime}\}\rho (t+\tau)
\displaystyle \longrightarrow
_{r \to -\infty} T exp \{i \int_{t}^{t+\tau} L(t^{\prime})dt^{\prime}\}
\rho_q (t+\tau). \eqno(3.7)
$$

Thus, the introduction of infinitively small source into Liuville equation
is the convenient method of selection of delayed solution  of this equation
that depends on $\gamma_{\alpha}(t)$ and satisfies the boundary condition in
(3.7).

In the Appendix the equivalence of the boundary condition (3.7) and the
mixing operations are shown:

$$
exp(iL_{0}\tau) \rho (t) \displaystyle \longrightarrow_
{\tau \to -\infty}
exp (iL_{0}\tau) \rho_q (\gamma (t)), \eqno(3.8)
$$
that is used in the method [3,9].

This fact is very remarkable the more so that we have the significant
difference in conditions (3.7) and (3.8). If in (3.7) evolution of the
system is considered from moment $t$ to moment $t+\tau$ along the
phase trajectory, i.e. with the whole Hamiltonian $H_t$ during which
are changed both the dynamical variables (multiplier
 $T\exp\{i\int_{t}^{t+\tau}L(t^{\prime}dt^{\prime}\}$),
and the macroscopic parameters $(\gamma_{\alpha}(t) \rightarrow \gamma_
{\alpha}(t+\tau))$, in (3.7) the mixing operation describing a free
evolution of the system takes place at fixed values of
$\gamma_{\alpha}(t)$.

The solution of Eq. (3.1) has the following form [6,8]:

$$
\rho (t) = \rho_q (t) - \int_{-\infty}^{0} d\tau e^{\epsilon \tau}
T exp\{i\int_{t}^{t+\tau} L(t^{\prime})dt^{\prime}\}\biggl(\frac
{\partial \rho_q (t+\tau)}{\partial \tau}\biggr) + iL(t+\tau) \rho_q (t+\tau).
\eqno(3.9)
$$

The use of explicit form of NSO (3.9) in the Appendix is difficult
because of its complex structure. In this respect the integral equation
for $\rho (t)$ [10], derived in the absence of variable fields for
 $\hat {\gamma_{\alpha}}$ operators satisfying (2.2) is much easier:

$$
\rho (t) = \rho_q (t) - \int_{-\infty}^{0} d\tau e^{(\epsilon+iL_0)\tau}
(1 - P_R (t+\tau)) iL_v (t+\tau) \rho (t+\tau), \eqno(3.10)
$$
where
$$
P_R (t)A = \sum_{\alpha} \frac{\partial \rho_q (t)}
{\partial \gamma_{\alpha} (t)}
Sp(\hat {\gamma_{\alpha}}A) -
$$

is the Robertson projection operator [11]. According to (3.2),
$P_R (t)A = P(t)A$, if $SpA=0$.

At $\epsilon = 0$ Eq. (3.10) coincides with the equation obtained by
the other method [12] when studying the system evolution under the
influence of external fields variables assuming that at
 $\tau \to -\infty$ the equilibrium condition of the system was disturbed
by adiabatic application of variable field.

In (3.9) and (3.10) $\rho (t)$ is non-Markov functional of
$\gamma_{\alpha} (t)$ as it depends on all values of  $\gamma_{\alpha}
(t^{\prime})$ á $t^{\prime} \leq t.$ As a consequence, the non-Markov
integral-differential generalized kinetic equations are obtained even in
the absence of high-frequency fields $(H_F (t) = 0)$ [2]:

$$
\frac{\partial \gamma_{\alpha} (t)}{\partial t} = Sp \hat {\gamma_{\alpha}}
\frac {\partial \rho (t)}{\partial t} = \\
Sp \rho (t) iL(t)
\hat {\gamma_{\alpha}} \equiv {L_{\alpha}}(t), \eqno(3.11)
$$
where ${L}_{\alpha}(t)$ is the collision integral.

On the other hand, in accordance with the requirement of the completeness
of the reduced description of irreversible processes
a nonequilibrium condition
and hence $\rho (t)$ at some moment of $t$ (from macroscopic point of view)
i.e. for $t \gg \tau_0$, should be determined by giving
$\gamma_{\alpha}(t)$ parameters at the same moment of time.
In particular problems the memory seems to be insignificant due to the
use of operator with (2.2) and (2.3) properties as
$\hat {\gamma_{\alpha}}$ operators and to the weak interactions
$V_t$ [4]. In stationary case $V_t = V \not = f(t),$ as it is shown in [13],
one can, in principle, obtain Markov presentation of NSO in any finite
order with respect to $V$  by reconstruction of a number of perturbation
theories with respect to weak interaction $V$ [10].

Taking into account that in widely and successfully used methods of GKE
construction the Markov character of NSO
$\rho (t)$ dependence on $\gamma_{\alpha} (t)$  is supposed from the
very beginning [1,3] it is more natural to formulate the Markov form
of NSO II method [6] in general form without using a definite order of
perturbation theory for $V_t$. The limitation of the class of Eq. (3.1)
solutions to Markov type eliminates the difficulties in solution of
non-Markov equations (3.10) in high orders of perturbation theory with
respect to $V_t$.
\newpage
\begin{center}
{4. Conversion of NSO to Markov form }
\end{center}

Let us introduce the operator of shift in time

$$
exp \{\tau \frac{\partial}{\partial \tau}\}\varphi (t) \equiv \\
\sum_{n=0}^{\infty} \frac {\tau^n}{n!} \frac {\partial^n}{\partial t^n}
\varphi (t) = \varphi (t+\tau).
$$
Then Eq. (3.10) has the following form:

$$
\rho (t) = \rho_q (t) - \int_{-\infty}^{0} d\tau e^{(\epsilon+iL_0)\tau}
e^{\tau\frac{\partial}{\partial \tau}}\{(1 - P_R (t))iL_v (t) \rho (t)\}.
\eqno(4.1)
$$
Let us proceed from the superposition

$$
\rho (t) \equiv \rho (\gamma (t),t); \qquad t \gg \tau_0, \eqno(4.2)
$$
assuming that $\rho (t)$ at $t \gg \tau_0$ depends on $t$ not only through
$\gamma_{\alpha}(t)$ value at the same moment of $t$, but by
$V_t$.

Note, that the time dependence of NSO just of such kind is obtained
as a result of direct reconstruction of a number of perturbation theories
[13], thus the supposition (4.2) can be considered as a mathematical
formulation of the possibility to construct Markov with respect to
$\gamma_{\alpha}(t)$ NSO expansion into series with respect to
$V_t$ perturbation. In this case the effect of $\partial /\partial t$
operator on arbitrary Markov type $f(\gamma (t),t)$ functional of
$\gamma_{\alpha}(t)$ is determined by the relation:

$$
\frac{\partial}{\partial t}f(\gamma (t),t) = \\
\biggl (\sum_{\alpha}
\frac{\partial \gamma_{\alpha} (t)}{\partial t} \frac{\partial}
{\partial \gamma_{\alpha} (t)} + \\
\frac {\tilde {\partial}}{\partial t}\biggr)
f(\gamma (t),t) \equiv ({D}(\gamma (t),t) + \frac{\tilde {\partial}}
{\partial t}) f(\gamma (t),t), \eqno(4.3)
$$
where $\frac {\tilde {\partial}}{\partial t}$ means the partial derivative
with respect to real time included in $f(\gamma (t),t)$ functional.

Differentiating (3.4) with respect to time and using (3.1), (2.2)
and (3.5), we obtain GKE

$$
\frac {\partial \gamma_{\alpha}(t)}{\partial t} = i\sum_{\beta}
{a}_{\alpha
\beta} \gamma_{\beta}(t) + $$
$$Sp (\rho (\gamma (t),t)iL_v \hat {\gamma_{\alpha}})
\equiv L_{\alpha} (\gamma (t),t), \eqno(4.4)
$$
where $L_{\alpha}$ is the collision integral of Markov type with respect to
$\gamma_{\alpha}(t)$.

According to (4.3) we can write

$$
D (\gamma (t),t) = \sum_{\alpha} L_{\alpha}(\gamma,t) \frac
{\partial}{\partial \gamma_{\alpha}} = \\
D_0 (\gamma) + D_v (\gamma,t);
$$
where
$$
D_0 (\gamma) = i \sum_{\alpha,\beta} a_{\alpha \beta} \gamma_{\beta}
\frac{\partial}{\partial \gamma_{\alpha}}; $$
$$D_v = \sum_{\alpha} Sp(\rho (\gamma,t) iL_v \gamma_{\alpha})
\frac{\partial}{\partial \gamma_{\alpha}}. \eqno(4.5)
$$

Introducing the symbols:

$$
B \equiv \rho (\gamma,t) - \rho_q (\gamma);$$
$$A \equiv - (1 - P_R (\gamma))iL_v (t) \rho (\gamma,t); $$
$$H \equiv \frac{1}{i}(\epsilon + iL_0 + D_0 (\gamma)); $$
$$U \equiv \frac{1}{i} (D_v (\gamma,t) + \frac{\tilde {\partial}}{\partial t},
\eqno(4.6)
$$
in view of Eqs. (4.2), (4.3), (4.5) and the rearrangement ability of
$(\epsilon + iL_0)$ and $D (\gamma,t)$, operators, Eq. (4.1)  can be
written as:

$$
B = \int_{-\infty}^{0} d\tau e^{i(H+U)\tau} A. \eqno(4.7)
$$

Eq. (4.7) is equivalent to integral equation [3]

$$
B = \int_{-\infty}^{0} d\tau e^{iH\tau} (A - iUB). \eqno(4.8)
$$

In previous notations, taking into account that  $exp(\tau D_0 (\gamma))$
operator describes free evolution of $(V_t = 0)$ from $t$ moment to
$t + \tau$ moment at which according to (4.4) $\gamma_{\alpha}(t) \to
\gamma_{\alpha}(t+\tau) = (e^{ia \tau}\gamma (t))_{\alpha} =
e^{\tau D_0 (\gamma)} \gamma_{\alpha} (t)$ Eq. (4.8)  takes the
following form:

$$
\rho (\gamma,t) = \rho_q (t) - \\
\int_{-\infty}^{0} d\tau e^{(\epsilon + iL_0)\tau}
\{iL_v (t) \rho (\gamma,t) + $$
$$+ \sum_{\alpha} \frac{\partial \rho (\gamma,t)}
{\partial \gamma_{\alpha}} Sp (\rho (\gamma,t)iL_v (t)\gamma_{\alpha}) +
\frac{\tilde {\partial}}
{\partial t}\rho(\gamma,t)\}_{\gamma \rightarrow e^{ia \tau}
\gamma}, \eqno(4.9)
$$
where the equality

$$
D_v (\gamma,t) \rho_q (\gamma) = \sum_{\alpha} \frac{\partial
{\rho_q (\gamma)}}{\partial {\gamma_{\alpha}}} Sp \{\rho (\gamma)iL_v (t)
\hat {\gamma_{\alpha}}\} = \\
- P_R (\gamma)iL_v (t) \rho (\gamma,t).
$$
is used.

Similarly, instead of (4.6) we use the notations (symbols):

$$
B \equiv \rho (\gamma,t) - \rho_q (\gamma);$$
$$A \equiv -(1 - P_R (\gamma))iL_v (t) \rho (\gamma,t);$$
$$H = \frac{1}{i} (\epsilon + iL_0 + D_0 (\gamma) + \frac{\tilde {\partial}}
{\partial t});$$
$$U = \frac{1}{i} D_v (\gamma,t), \eqno(4.10)
$$
we obtain:

$$
\rho (\gamma,t) = \rho_q (\gamma) - \\
\int_{-\infty}^{0} d\tau e^{(\epsilon+
iL_0)\tau} \{iL_v (t + \tau) \rho (\gamma,t + \tau) + $$
$$+ \sum_{\alpha} \frac{\partial \rho (\gamma,t+\tau)}
{\partial \gamma_{\alpha}}
Sp \rho (\gamma,t + \tau)il_v (t + \tau) \hat {\gamma_{\alpha}}\}_{\gamma
\rightarrow e^{i\alpha \tau}\gamma}. \eqno(4.11)
$$

It is easily seen that Exps.  (4.9) and (4.11) are strictly equivalent and
automatically satisfy the matching conditions (3.4).

When $\epsilon = 0$, (4.9) coincides with the equation obtained directly
from Liuville equations using boundary conditions (3.8) in low-frequency
region $\omega \tau_0 \ll 1$ of variable field [3], and Eq. (4.11)
coincides with the equation derived by similar method for arbitrary
values of $\omega \tau_0$ parameter [14]. In stationary condition
$({H}_{F}(t) = 0)$, when NSO
$\rho (t)$ does not depend explicitly on time at  $\epsilon = 0$
Eqs. (4.9) and (4.11) coincides with the basic equation of the method
[9]. The same result assuming
$\rho (t) = \rho (\gamma (t))$ at $t \gg \tau_0$
can be obtained directly from Eq.  (3.1) [15].

In our method Eqs. (3.10), (4.9) and (4.11)
are similarly applicable at arbitrary values of
 $\omega_0 \tau$ parameter. Nevertheless, in the low-frequency region
when $\omega \tau_0 \ll 1$ and correspondingly
$\frac{\tilde {\partial} \rho (\gamma,t)}{\partial t} \sim \omega \tau_0$
term is small it is convenient to use Eq. (4.9) and at high frequencies
$\omega \tau_0 \geq 1$ is preferable to use Eq. (4.11).

In general case at arbitrary $\omega \tau_0$ GKE (4.4) are differential with
respect to $t$, i.e. belong to Markov type, though they contain the
memory with of respect to variable field. At $\omega \tau_0 \ll 1$ the
memory with respect to field disappears and the evolution of the system in
time in the variable field is strictly Markov type.

\begin{center}
{5. Discussion of results}
\end{center}

Let us discuss the conditions of applicability of the obtained equations.
According to (3.6)
$\rho (-\infty) = \rho_q (-\infty) = \rho_0$ ($\rho_0$ -
is the equilibrium statistical operator of the system).
To form the quasiequilibrium state
$\sim \tau_0$ time is required. Therefore NSO  $\rho (\gamma,t)$ and
Eqs. (3.10), (4.9),
(4.11) describe the system at  $t \gg \tau_0$.

Let us assume that in remote part the system was in equilibrium
$(\rho (-\infty) = \rho_0)$ and then the equilibrium was disturbed
by adiabatic application of variable fields. As is shown in [3,12]
at such boundary condition the limitation of $t \gg \tau_0$
is removed and Eq. (3.10) and hence Eqs.
(4.9) and (4.11) following from it describe the system adequately all
$t$ time.

As for the frequencies $\omega$ and amplitudes $\omega_1$ of variable fields
Eqs.(3.10), (4.9) and (4.11) are applicable at arbitrary frequencies and
at the amplitudes for which
$H_F (t) \ll H_0$.

Macroscopic parameters $\gamma_{\alpha}(t)$ usually are introduced for
description
of nonequilibrium states of systems, the Hamiltonian of which does not depend
on time. At the action of variable fields the time-dependence of
Hamiltonian is eliminated by the transition into rotating coordinate
system [16] or by considering the fields from quantum point of view
 [17] and only after this $\gamma_{\alpha}(t)$ is introduced.
When  $\hat {\gamma_{\alpha}}$ operators commutate with  $H_0$ i.e.
$a_{\alpha \beta} \equiv 0$ such approach preserves the common thermodynamic
sense of  $\gamma_{\alpha}(t)$ values and is the most consistent. But
such approach is not always applicable, therefore one should discuss
the problem of possibility to use
 $\gamma_{\alpha}(t)$ parameters in general case.

First, let us assume that $a_{\alpha\beta} \equiv 0$. In the low-frequency
region $\gamma_{\alpha}(t)$ parameters, being the slowly changing
time functions preserve the common thermodynamic sense. At each instant
of $t$ time the system is "adjusted" to the instant value of
Hamiltonian $H_t$ (2.1).
Then the whole difference from the stationary case is reduced
only to that for  $t \gg \tau_0$ the statistical operator
$\rho (t)$ depends on  $t$ not only through
$\gamma_{\alpha}(t)$, but explicitly through the variable field
 $F(t)$ and its derivatives $\dot F(t), \ddot F(t), \ldots$ [3].
As it mentioned in [3], for $\omega \tau_0 \ll 1$ the kinetic processes
are in fact the Markov type and (4.9) is obtained from (3.10) by
direct expansion of $V_{t+\tau}$ and $\gamma_{\alpha}(t+\tau)$
values into power series $\tau$. In our aforementioned approach,
Eqs.(3.10) and (4.9) are strictly equivalent for any values of
 $\omega \tau_0$ parameter.

In the high-frequency region $\omega\tau_0 \geq 1$ the non-Markov
character of kinetic processes, taking place in the system under
the action of variable fields will manifest itself.
In this case the expansion in terms of $\omega\tau_0$
parameter seems impossible. $\gamma_{\alpha}(t)$ parameters have no
longer the direct thermodynamic sense as they contain the rapid,
but low-amplitude (because $H_F (t) \ll H_0$)
dynamic change induced by high-frequency field. Nevertheless,
 $\gamma_{\alpha}(t)$ parameters can be still used for description of
quasiequilibrium states, as, according to (3.2), owing to the
"mixing" action of $H_0$, the system will have time to be "adjusted" to
the instant nonequilibrium values of  $\gamma_{\alpha}(t)$
parameters.

When $a_{\alpha\beta} \not\equiv 0$ and $V_t = 0$,
$\gamma_{\alpha}(t)$ values in accordance with (4.4) are changed by the
law $\gamma (t) = e^{i a t} \gamma (0)$ with arbitrary frequencies
determined by eigenvalues of  $(e^{i a t})_{\alpha\beta}$ matrics.
Therefore, $\gamma_{\alpha}(t)$ are not "slow" thermodynamic values even
in the absence of variable fields. However, it is remarkable that in
the general case
$ a_{\alpha\beta} \not\equiv 0, \omega\tau_0 \geq 1$
the "adjusting" of the system to the instant values $\gamma_{\alpha}(t)$
takes place and the dependence of $\rho (\gamma (t),t)$ on
$\gamma_{\alpha}(t)$, as well as Eq.
(4.11), are of Markov character with respect to $\gamma_{\alpha}(t)$,
though they contain the memory with respect to variable field.

\begin{center}
{Conclusion}
\end{center}

The carried out study shows the full equivalence of the boundary condition
(3.7) and mixing operation (3.8). Thus, we prove the strict equivalence
of the method
 [3,9,12] to the variety of NSO method [6]
based on Liuville equation with infinitively small source
(3.1) and $\hat {\gamma_{\alpha}}$ operators with  (2.2) properties.

The main conclusion of the work is that the method NSO II [6]
can be formulated in Markov form with respect to $\gamma_{\alpha}(t)$
parameters even in the presence of high-frequency fields. As a result,
the use of high order perturbation theory with respect to small
interactions in the system is considerably facilitated.

\newpage

\begin{center}
{Appendix}
\end{center}

Let us show the equivalence of boundary condition (3.7)
and mixing operation (3.8).

Liuville equation (3.5) can be written as

$$
\biggl (\frac{\partial}{\partial t} + iL_0\biggr) \delta \rho (t) = -\\
(1 - P_R (t)) iL_v (t) \rho (t); \delta \rho = \rho - \rho_q. \eqno(A1)
$$
Let us multiply (A1) by integrating factor $exp (iL_0 t)$  and integrate
it from
 $-\infty$ to $t$.

Note, that the solution of Eq. (3.1) has the  [6,8]:

$$
\rho (t) = \rho_q (t) - \int_{-\infty}^{0} d\tau e^{\epsilon \tau} T
\exp \biggl \{i \int_{t}^{t+\tau} L(t^{\prime})dt^{\prime} \biggr \}
\biggl(\frac{\partial \rho_q (t+\tau)}{\partial \tau} + iL(\tau)
\rho_q (t+\tau)\biggr) = $$
$$= \rho_q (t) - \int_{-\infty}^{0} d\tau
e^{\epsilon \tau} T(t,t+\tau) (1 - P_R (t+\tau))iL(t+\tau)\rho_q (t+\tau),
\eqno(A2)
$$
where
$$
T(t,t_1) = T \exp \biggl\{- \int_{t_1}^{t} (1 - P_R (t^{\prime}))
iL(t^{\prime}) dt^{\prime} \biggr\},
$$
and $P_R(t)$ is determined in the main text (see (3.10)).

Using (2.2) and the relation [3,8],
$$
e^{-iL_0 \tau} \rho_q (\gamma (t)) = \rho_q (e2^{i\alpha \tau} \gamma (t)),
\eqno(A3)
$$
we obtain:
$$
e^{iL_0t} \delta \rho (t) = - \int_{-\infty}^{t} dt^{\prime}
e^{iL_0 t^{\prime}}
(1 - P_R (t^{\prime}))iL_v (t^{\prime}) \rho (t^{\prime}). \eqno(A4)
$$
Taking into account that according to (3.8),
$$
\displaystyle\lim_{\tau\to -\infty} e^{iL_0 \tau} (\rho (\tau) -
\rho_q (\tau)) = 0,
$$
after simple transformation of (A4) we have
$$
\rho (t) = \rho_q (t) - \int_{-\infty}^{0} d\tau e^{iL_0 \tau}
(1 - P_R (t+\tau) iL_v (t+\tau) \rho (t+\tau). \eqno(A5)
$$
(A5) coincides with (3.10) at $\epsilon = 0$. The forming factor
$e^{\epsilon \tau}$ does not affect the integral convergence [13,15]
included in Eq. (3.10) and Eqs. (3.6), (3.8) and (3.6), (3.7) are fully
equivalent as they result in coinciding integral equations (3.10)
and  (A5).

\end{document}